# Observation of Target Electron Momentum Effects in Single-Arm Møller Polarimetry⋆

M. Swartz,(5) H.R. Band,(6) F.J. Decker,(5) P. Emma,(5) M.J. Fero,(3)
R. Frey,(4) R. King,(5) A. Lath,(3) T. Limberg,(5) R. Prepost,(6)
P.C. Rowson,(1) B.A. Schumm,(2) M. Woods,(5) and M. Zolotorev (5)

(1) *Columbia University, New York, New York 10027 USA*
(2) *Lawrence Berkeley Laboratory, University of California, Berkeley, California 94720 USA*
(3) *Massachusetts Institute of Technology, Cambridge, Massachusetts 02139 USA*
(4) *University of Oregon, Eugene, Oregon 97403 USA*
(5) *Stanford Linear Accelerator Center, Stanford University, Stanford, California 94309 USA*
(6) *University of Wisconsin, Madison, Wisconsin 53706 USA*

## ABSTRACT

In 1992, L.G. Levchuk noted that the asymmetries measured in Møller scattering polarimeters could be significantly affected by the intrinsic momenta of the target electrons. This effect is largest in devices with very small acceptance or very high resolution in laboratory scattering angle. We use a high resolution polarimeter in the linac of the polarized SLAC Linear Collider to study this effect. We observe that the inclusion of the effect alters the measured beam polarization by −14% of itself and produces a result that is consistent with measurements from a Compton polarimeter. Additionally, the inclusion of the effect is necessary to correctly simulate the observed shape of the two-body elastic scattering peak.

Submitted to *Nuclear Instruments and Methods*

⋆ Work supported by the Department of Energy, contract DE–AC03–76SF00515.

## 1. Introduction

In 1992, L.G. Levchuk noted that the asymmetries measured in Møller scattering polarimeters could be significantly affected by the intrinsic momenta of the target electrons.[1] He estimated that the asymmetries measured by several polarimeters at the MIT-Bates laboratory would be increased by 5-10% where the exact value depends upon the acceptance and resolution in laboratory scattering angle. He also predicted that this effect would be small in the large acceptance SLAC polarimeters. We note that although the SLAC polarimeters do have large acceptance, some have high angular resolution and should be quite sensitive to effects caused by the intrinsic momenta of the target electrons.

The SLAC Linear Collider (SLC) provides an ideal environment in which to study target momentum effects. It includes: a high energy electron beam of very small emittance and large polarization, a Møller polarimeter with high angular resolution, and a precise Compton polarimeter to monitor the beam polarization. This paper describes a study of the effects of intrinsic target momentum upon the angular size of the two-body elastic peak and upon the magnitude and angular shape of the measured Møller asymmetry.

## 2. Møller Polarimetry

The scattering of a polarized electron beam from the polarized electrons in a magnetized target is a common technique for the measurement of the beam polarization. Assuming that the square of center-of-mass (cm) energy of the two-electron system, $s$, is much larger than the square of the electron mass, the tree-level differential cross section for this process in the cm-frame can be expressed as follows,

$$\frac{d\sigma_0}{d\Omega}(s) = \frac{\alpha^2}{s}\frac{(3+\cos^2\hat{\theta})^2}{\sin^4\hat{\theta}}\Big\{1 - \mathcal{P}_z^B\mathcal{P}_z^T A_z(\hat{\theta}) - \mathcal{P}_t^B\mathcal{P}_t^T A_t(\hat{\theta})\cos(2\phi-\phi_B-\phi_T)\Big\}, \quad (1)$$

where: $\alpha$ is the fine structure constant; $\hat{\theta}$ is the cm-frame scattering angle; $\phi$ is the azimuth of the scattered electron (the definition of $\phi = 0$ is arbitrary); $\mathcal{P}_z^B$, $\mathcal{P}_z^T$ are the longitudinal polarizations of the beam and target, respectively; $\mathcal{P}_t^B$, $\mathcal{P}_t^T$ are the transverse polarizations of the beam and target, respectively; $\phi_B$, $\phi_T$ are the azimuths of the transverse polarization vectors; and $A_z(\hat{\theta})$ and $A_t(\hat{\theta})$ are the longitudinal and transverse asymmetry functions which are defined as

$$\begin{aligned} A_z(\hat{\theta}) &= \frac{(7+\cos^2\hat{\theta})\sin^2\hat{\theta}}{(3+\cos^2\hat{\theta})^2} \\ A_t(\hat{\theta}) &= \frac{\sin^4\hat{\theta}}{(3+\cos^2\hat{\theta})^2}. \end{aligned} \tag{2}$$

The asymmetry functions are maximal at 90° scattering ($A_z(90°) = 7/9$, $A_t(90°) = 1/9$) and approach zero in the forward and backward directions.

In order to determine the beam polarization, the rate of electrons scattered into some solid angle $d\Omega$ is measured for a fixed relative orientation of the beam and target polarization vectors $R(\mathcal{P}^B\mathcal{P}^T)$ and with one polarization vector inverted $R(-\mathcal{P}^B\mathcal{P}^T)$. The asymmetry formed from these rates $A_R$ is then simply related to the beam and target polarizations:

$$\begin{aligned} A_R &\equiv \frac{R(\mathcal{P}^B\mathcal{P}^T) - R(-\mathcal{P}^B\mathcal{P}^T)}{R(\mathcal{P}^B\mathcal{P}^T) + R(-\mathcal{P}^B\mathcal{P}^T)} \\ &= -\mathcal{P}_z^B\mathcal{P}_z^T A_Z(\hat{\theta}) - \mathcal{P}_t^B\mathcal{P}_t^T A_t(\hat{\theta})\cos(2\phi - \phi_B - \phi_T). \end{aligned} \tag{3}$$

The beam polarization is extracted from the measured value of $A_R$, the measured target polarization, and the theoretical asymmetry functions.

The actual polarization measurement is performed in the laboratory frame. The Lorentz transformation is normally performed with the assumption that the target electron is a free particle at rest in the laboratory frame. In this approximation, the square of the center-of-mass energy $s_0$ is given by the following simple expression,

$$s_0 = 2p_b m_e, \tag{4}$$

where $p_b$ is the beam momentum and $m_e$ is the electron mass. The relationship between the center-of-mass scattering angle and the laboratory momentum of the scattered electron, $p'$, is given by the following expression,

$$p' = \frac{p_b}{2}(1 + \cos\hat{\theta}). \tag{5}$$

In small angle approximation, the laboratory scattering angle $\theta$ is given as follows:

$$\theta^2 = \frac{1}{p_b p'} \cdot \frac{s_0}{2}(1 - \cos\hat{\theta}) = 2m_e \left(\frac{1}{p'} - \frac{1}{p_b}\right). \tag{6}$$

Equation (6) is the basis of all single-arm Møller polarimetry. The correlation between $\theta$ and $p'$ is used to identify the electron-electron elastic scattering signal. A schematic diagram of a single-arm Møller polarimeter is shown in Figure 1. A narrow slit located downstream of the target defines the scattering plane. The scattered electrons are momentum analyzed by magnetic deflection in the plane that is perpendicular to the scattering plane. In the case that a dipole magnetic field is used, the elastically scattered electrons produce a parabolically-shaped line image on a downstream detector plane. In most polarimeters, the momentum acceptance is sufficiently small as compared with the angular acceptance that the accepted segment of the parabola is approximated well by straight line. A position sensitive detector is oriented so that it measures the number of incident electrons as a function of the coordinate that is perpendicular to the accepted line segment. Therefore, elastically scattered electrons appear as a narrow peak on the detector. Signal from various background sources does not prefer the region of the elastically-scattered peak and appears as a smooth distribution across the detector.

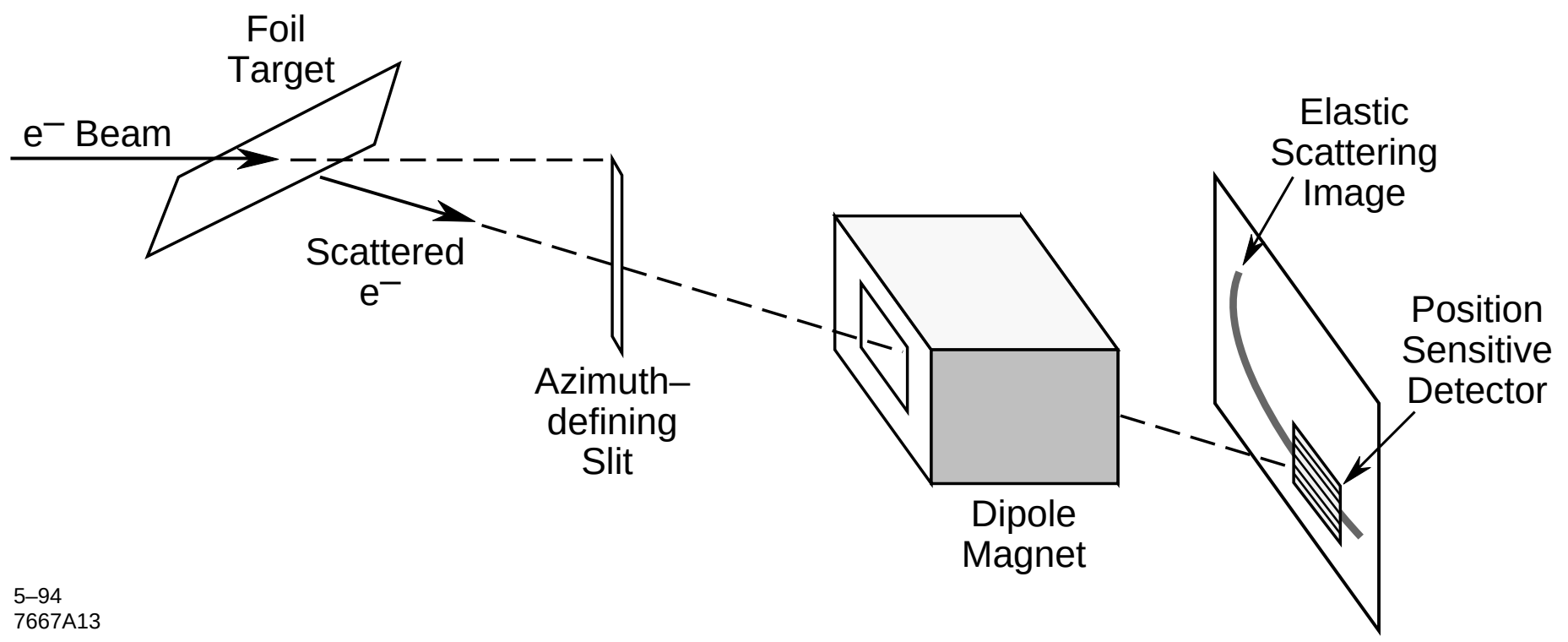

Figure 1. A schematic diagram of a single-arm Møller polarimeter.

### 2.1 The Levchuk Effect

The Levchuk Effect follows from the observation that the target electrons are not free particles at rest but are bound to atomic sites. The detailed kinematics of the scattering of a high energy electron from a bound state electron are discussed in Reference 1. In the high beam-energy limit, we can ignore the binding energy of the electron and the energy-momentum of the recoiling ion. To leading order, the square of the center-of-mass energy, $s_1$, is then given by the following expression,

$$s_1 = s_0 \left(1 - \frac{\vec{p_t} \cdot \hat{n}}{m_e}\right), \tag{7}$$

where $\vec{p_t}$ is momentum of the target particle, and $\hat{n}$ is the direction of the beam particle. Note that $s_0$ is smeared by a factor which ranges from $1 - p_t/m_e$ to $1 + p_t/m_e$ depending upon the target electron direction of motion. Since K-shell electrons can have momenta of order 100 KeV/c, this effect can be as large as 20%.

The presence of non-zero target particle momentum does not modify the relationship between the center-of-mass scattering angle (Møller asymmetry) and the laboratory momentum of the scattered electron because the $\sqrt{s_1}$ dependence of the Lorentz $\gamma$-factor cancels the dependence upon the center-of-mass energy scale,

$$p' = \frac{p_b}{\sqrt{s_1}} \cdot \frac{\sqrt{s_1}}{2}(1 + \cos\hat{\theta}) = \frac{p_b}{2}(1 + \cos\hat{\theta}). \tag{8}$$

However, the laboratory scattering angle is affected by the presence of non-zero target particle momentum,

$$\theta^2 = \frac{1}{p_b p'} \cdot \frac{s_1}{2}(1 - \cos\hat{\theta}) = 2m_e \left(\frac{1}{p'} - \frac{1}{p_b}\right) \cdot \left(1 - \frac{\vec{p}_t \cdot \hat{n}}{m_e}\right). \tag{9}$$

The laboratory scattering angle is smeared by the square root of the target-momentum-dependent factor that modifies the square of the center-of-mass energy.

Equation (9) is the basis of the Levchuk Effect. The presence of randomly-oriented, non-zero target electron momentum broadens the line image (in $\theta$-$1/p'$ space) at the detector plane. The degree of broadening is not uniform for all electrons in the target foil but depends upon the particular quantum state of the target electron. The targets used in most Møller polarimeters are composed of an iron-cobalt-vanadium alloy known as Vanadium-Permendur (49% Fe, 49% Co, 2% V). The K- and L-shell electrons in this material are unpolarized and have large mean momenta ($\sim$90 KeV/c and $\sim$30 KeV/c, respectively). The polarized electrons reside in the M-shells of the iron and cobalt atoms which along with the few N-shell electrons have smaller mean momenta ($\sim$10 KeV/c and $\sim$2 KeV/c, respectively). The images produced by elastic scattering from the unpolarized inner-shell electrons are therefore broader than those produced by scattering from the more highly polarized outer-shell electrons.

A simulation (described in Section 5.1) of this effect for the SLC Linac Møller Polarimeter is shown in Figure 2. The signal per target electron observed in each of the detector channels is shown for the K-, L-, M-, and N-shells of the iron atom. The net effect is to produce a nonuniformity in the observed scattering asymmetry as a function of detected coordinate. The asymmetry function is enhanced near the center of the peak and is depleted in the wings of the distribution. The resulting fractional effect upon the measured beam polarization depends upon the details of the analysis procedure but can be as large as 10-15%. Note that each of the signal peaks shown in Figure 2 has the same area. Therefore, the signal measured by a detector of large granularity or poor resolution is independent of the target electron momentum distribution and the effect is negligible.

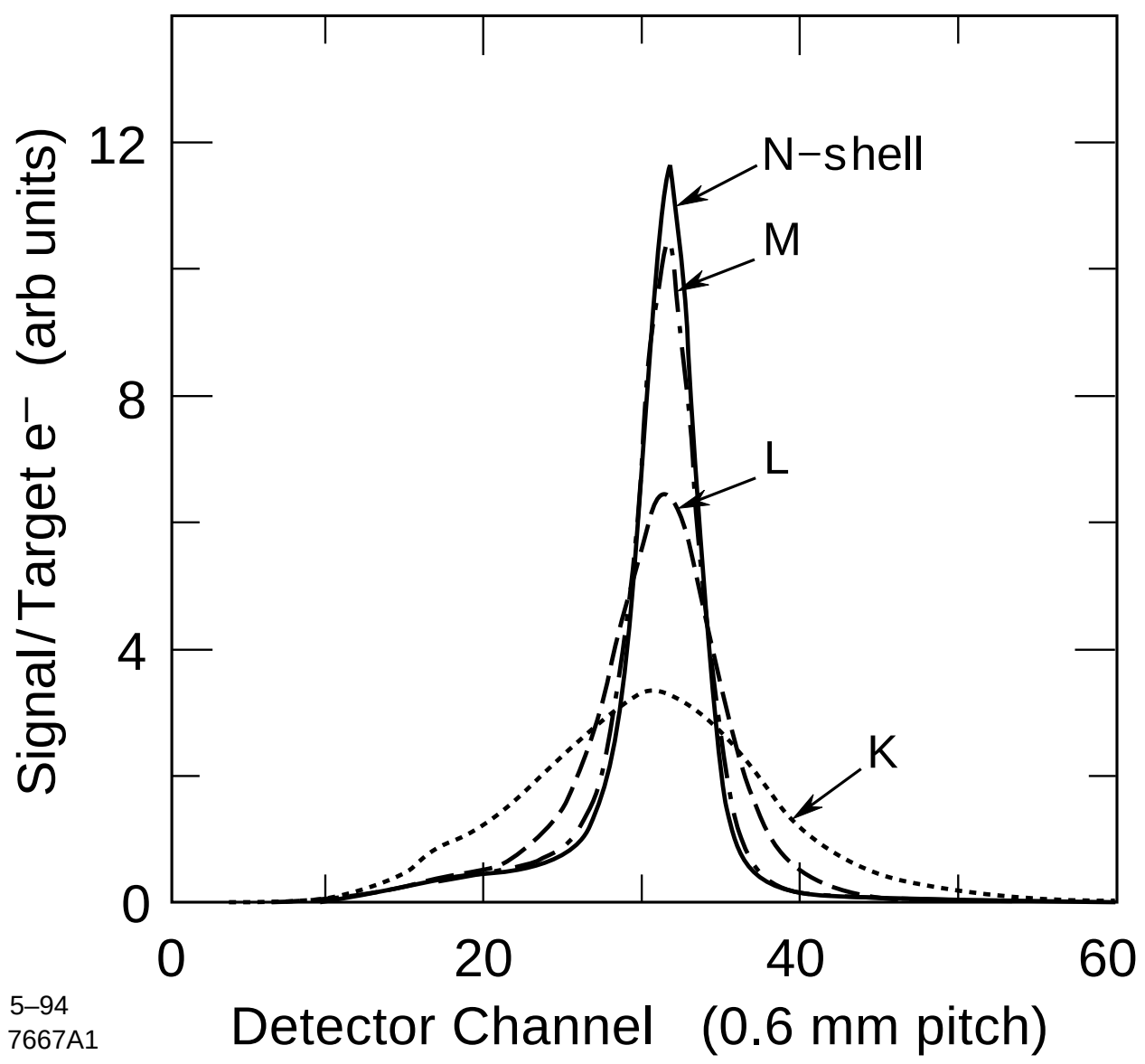

Figure 2. The simulated signal observed at the SLC Linac Møller detector per target electron for each of the atomic iron shells.

## 3. The Polarized SLC

A diagram of the polarized SLC is shown in Figure 3. Longitudinally polarized electrons are produced in the 120 kV Polarized Electron Source (PES) by photoemission from a strained-lattice GaAs cathode[2] illuminated by a pulsed Titanium-Sapphire laser[3] operating at a wavelength of 865 nm. The electron helicity is changed randomly on a pulse-to-pulse basis by changing the circular polarization of the laser beam. The PES produces 2 ns pulses of electrons which are compressed to 15 ps duration in several RF bunchers and are then accelerated to 1.19 GeV for storage in the North Damping Ring of the SLC. A system composed of the dipole magnets of the Linac-To-Ring transfer line and a superconducting solenoid magnet is used to rotate the longitudinal polarization of the beam into the vertical direction for storage in the damping ring. The Spin Rotation System[4] consisting of two superconducting solenoids and the dipole magnets of the Ring-To-Linac transfer line can be used to re-orient the polarization vector upon extraction from the damping ring. This system has the ability to provide nearly all polarization orientations in the linac.

Upon extraction from the damping ring, the polarized electron pulses are accelerated in the linac to 46.6 GeV. The SLC Møller Polarimeter is located in the beam switchyard between the linac and the beginning of the North Arc and is used for diagnostic purposes. Polarized electron pulses are then transported through the North Arc and Final Focus systems of the SLC to the interaction point (IP) of the machine. The North Arc is composed of 23 achromats, each of which consists of 20 combined function magnets. The average spin precession in each achromat is 1085° which is quite close to 1080° betatron phase advance caused by the same elements. The arc therefore operates near a spin resonance. In 1993, this feature was used to convert the final third of the arc into a spin rotator.[5] In normal operation, the solenoid-based spin rotation system is turned off and a vertically polarized electron beam is accelerated in the linac. A pair of large amplitude betatron oscillations in the final third of the arc is then used to rotate the polarization vector into the longitudinal direction at the SLC interaction point. The emission of synchrotron radiation in the arc reduces the energy of the beam to 45.65 GeV and slightly increases the energy spread of the transmitted beam (the RMS contribution of the arc is 0.06% which must be combined in quadrature with the 0.2% input energy spread). After passing through the interaction point, the longitudinal polarization of the electron beam ($\mathcal{P}_z^C$) is measured with a Compton polarimeter.

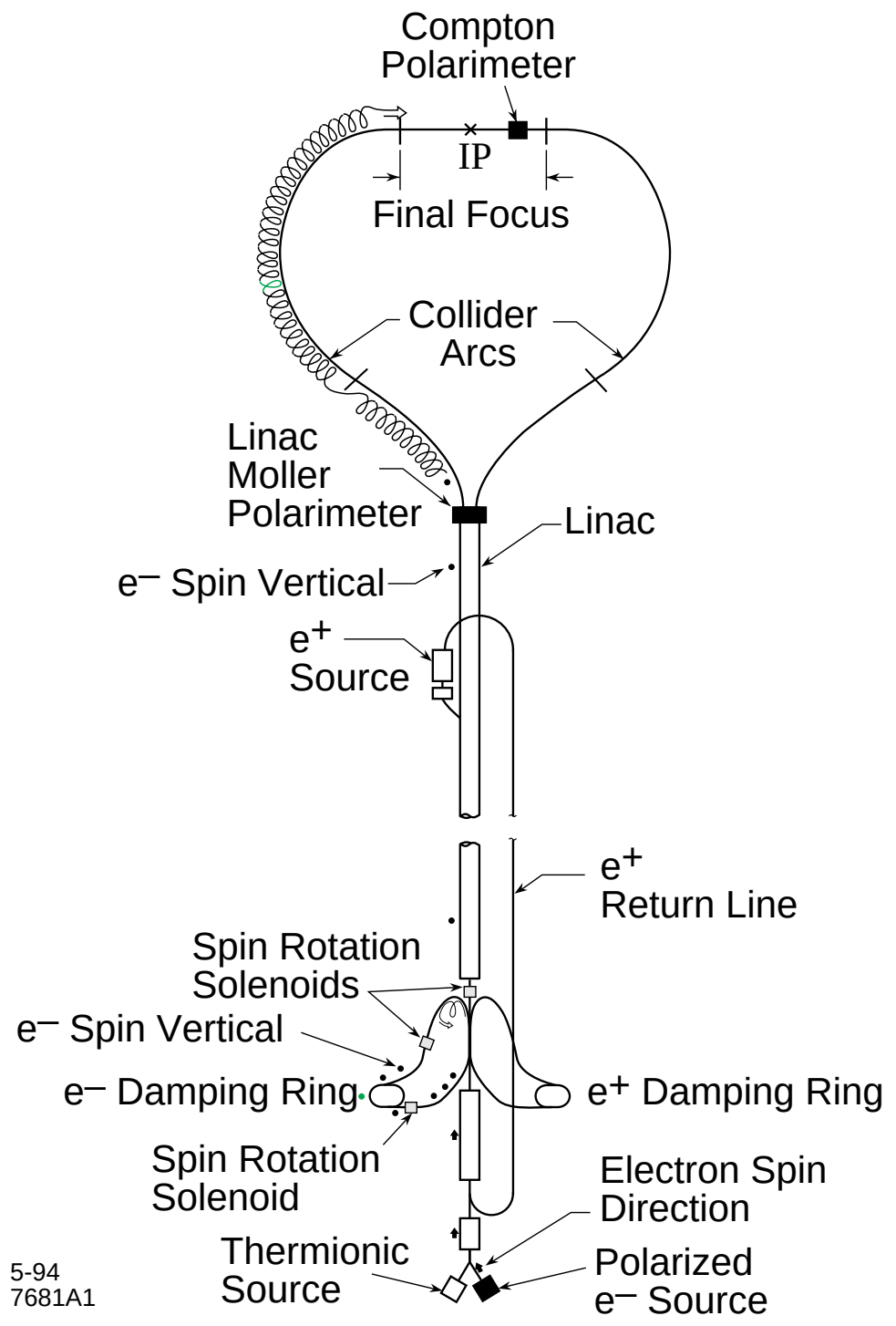

Figure 3. The polarized SLC. The electron spin direction is indicated by the double-arrow.

The beam is then transported through an extraction line to a beam dump.

### 3.1 The Compton Polarimeter

The Compton scattering polarimeter,[6] shown in Figure 4, is located 33 m downstream of the IP. After it has passed through the IP and before it is deflected by dipole magnets, the electron beam collides with a circularly polarized photon beam produced by a frequency-doubled Nd:YAG laser of wavelength 532 nm. The scattered and unscattered electrons remain unseparated until they pass through a pair of dipole magnets. The scattered electrons are dispersed horizontally and exit the vacuum system through a thin window. Multichannel Cherenkov and proportional tube detectors measure the momentum spectrum of the electrons in the interval from 17 to 30 GeV/c.

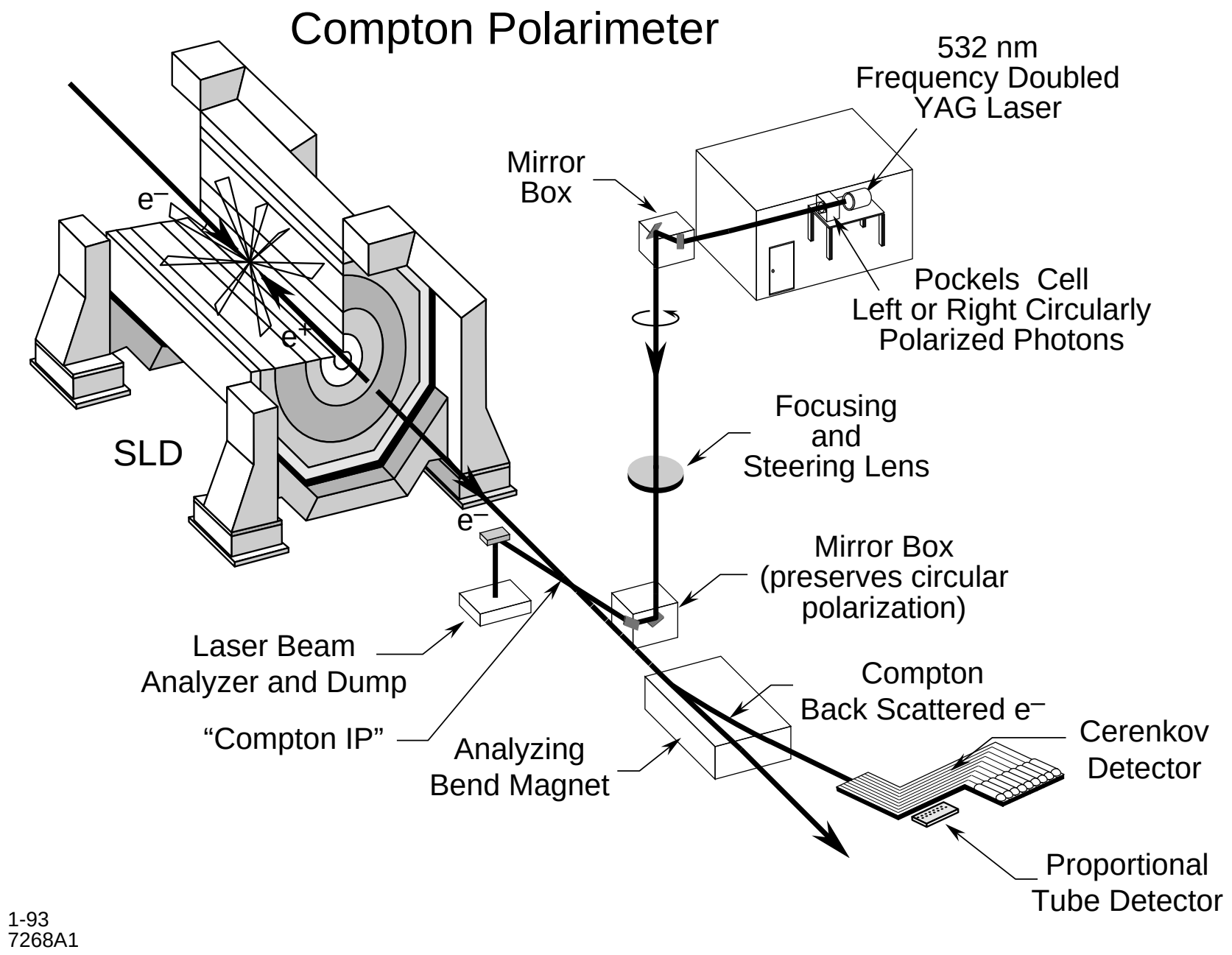

Figure 4. A schematic diagram of the SLC Compton Polarimeter.

The counting rates in each detector channel are measured for parallel and anti-parallel combinations of the photon and electron beam helicities. The asymmetry formed from these rates is equal to the product $\mathcal{P}_z^C \mathcal{P}_\gamma A(E)$ where $\mathcal{P}_\gamma$ is the circular polarization of the laser beam at the electron-photon crossing point and $A(E)$ is the theoretical asymmetry function (corrected for small detector acceptance and resolution effects) at the accepted energy $E$ of the scattered electrons[7]. The average channel-by-channel polarization asymmetry for a large sample of data is shown as a function of the mean accepted energy of each channel in Figure 5. The curve represents the product of $A(E)$ and a normalization factor $(\mathcal{P}_z^C \mathcal{P}_\gamma)$ that has been adjusted to achieve a best fit to the measurements. The laser polarization $\mathcal{P}_\gamma$ was maintained at 0.992±0.006 by continuously monitoring and correcting phase shifts in the laser transport system. The energy scale of the spectrometer is calibrated from measurements of the kinematic endpoint for Compton scattering (17.36 GeV) and the zero-asymmetry energy (25.15 GeV).

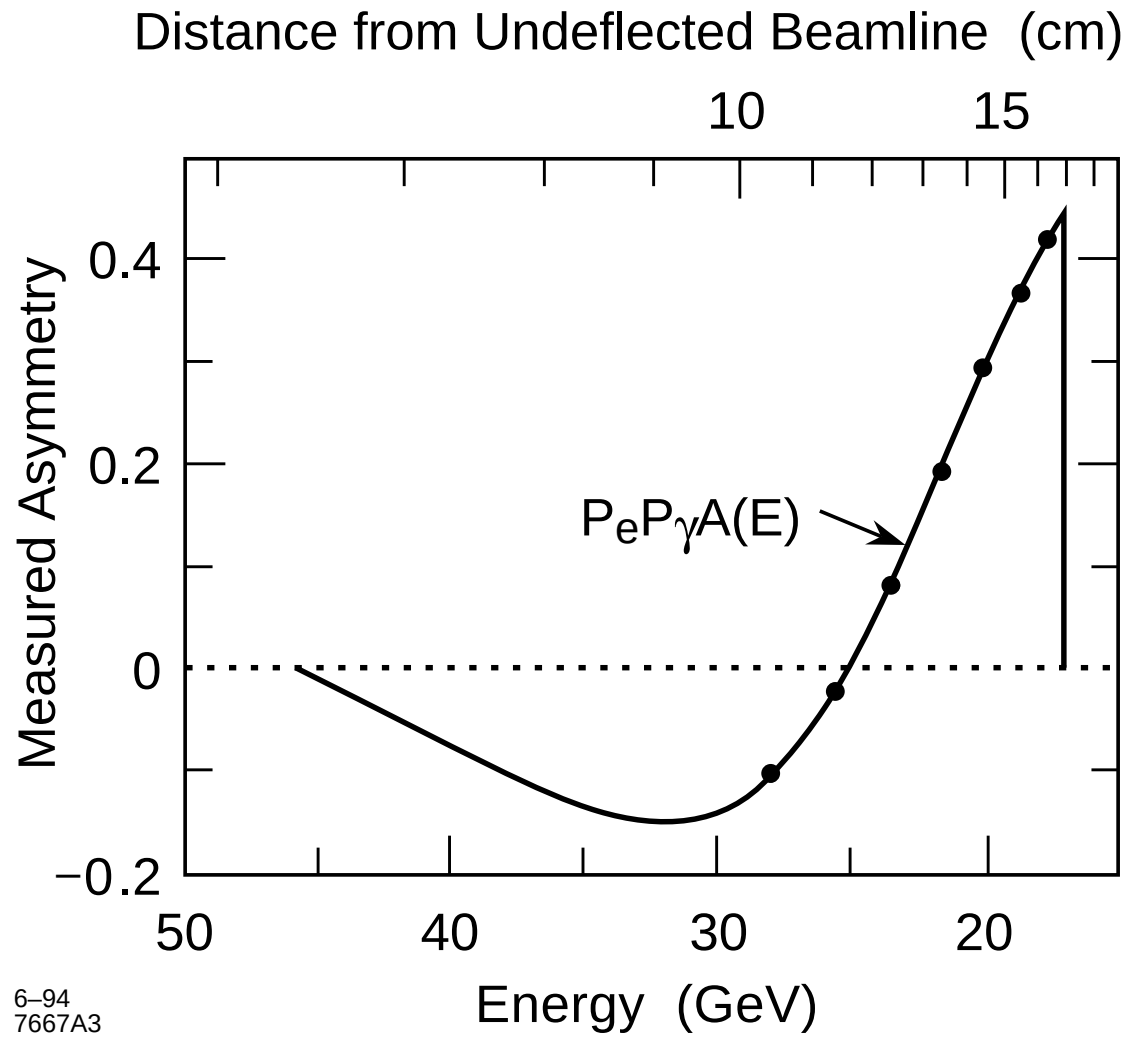

Figure 5. The average polarized Compton scattering asymmetry as measured by seven channels of the Cherenkov detector is plotted as a function of the mean accepted energy of each channel. The statistical errors are small as compared with the point size.

Polarimeter data are acquired continually for runs of approximately 3 minutes. For each run, $\mathcal{P}_z^C$ is determined from the observed asymmetry using the measured value of $\mathcal{P}_\gamma$ and the theoretical asymmetry function. The absolute statistical precision of each run is typically $\delta\mathcal{P}_z^C = 0.01$. The systematic uncertainties that affect the polarization measurement are summarized in Table 1. The total relative systematic uncertainty is estimated to be $\delta\mathcal{P}_z^C/\mathcal{P}_z^C = 1.1\%$.

**Table 1**. Systematic uncertainties that affect the Compton Polarimeter measurements.

| Systematic Uncertainty | $\delta\mathcal{P}_z^C/\mathcal{P}_z^C$ |
|---|---|
| Laser Polarization | 0.6% |
| Detector Linearity | 0.6% |
| Interchannel Consistency | 0.5% |
| Spectrometer Calibration | 0.4% |
| Electronic Noise | 0.2% |
| Total Uncertainty | 1.1% |

### 3.2 The SLC Linac Møller Polarimeter

The SLC Linac Møller Polarimeter is located in the beam switchyard of the linear accelerator complex. A schematic diagram of the polarimeter is shown in Figure 6. The 46.6 GeV electron beam is brought into collision with one of five insertable magnetized Vanadium-Permendur foils. Scattered electrons impinge upon an azimuth-defining collimator (labelled PC-0) located 4.10 m downstream of the target. The collimator accepts electrons that are scattered within ±75 mrad (azimuthal angle) of the downward vertical direction and have scattering angles between 5.9 and 8.4 mrad. The transmitted electrons are then deflected horizontally by a pair of dipole magnets at the entrance to the original PEP injection line. The bend angle of the central ray is 131 mrad and the effective bend center is located 4.21 m downstream of the collimator. A horizontal, momentum-defining slit is located 3.54 m downstream of the effective bend center. The width of the slit is adjusted to transmit electrons with momenta that are within ±3.1% of the 14.5 GeV/c central momentum. Finally, the transmitted electrons impinge upon a position sensitive detector located 1.36 m downstream of the momentum-defining slit. The detector consists of a two-radiation-length thick tungsten-lead radiator followed by a silicon strip detector. The detector has an active area of 56×38 mm consisting of 128 strips of 0.3 mm pitch. Since alternate strips are read-out via charge-sensitive preamplifiers and peak-sensing ADC's, the detector effectively has 64 channels of 0.6 mm pitch. The strip axis is rotated by 5.7° from the horizontal direction to account for the scattering angle-momentum correlation of the Møller image. The scattering angle resolution of the polarimeter is approximately 27 microradians.

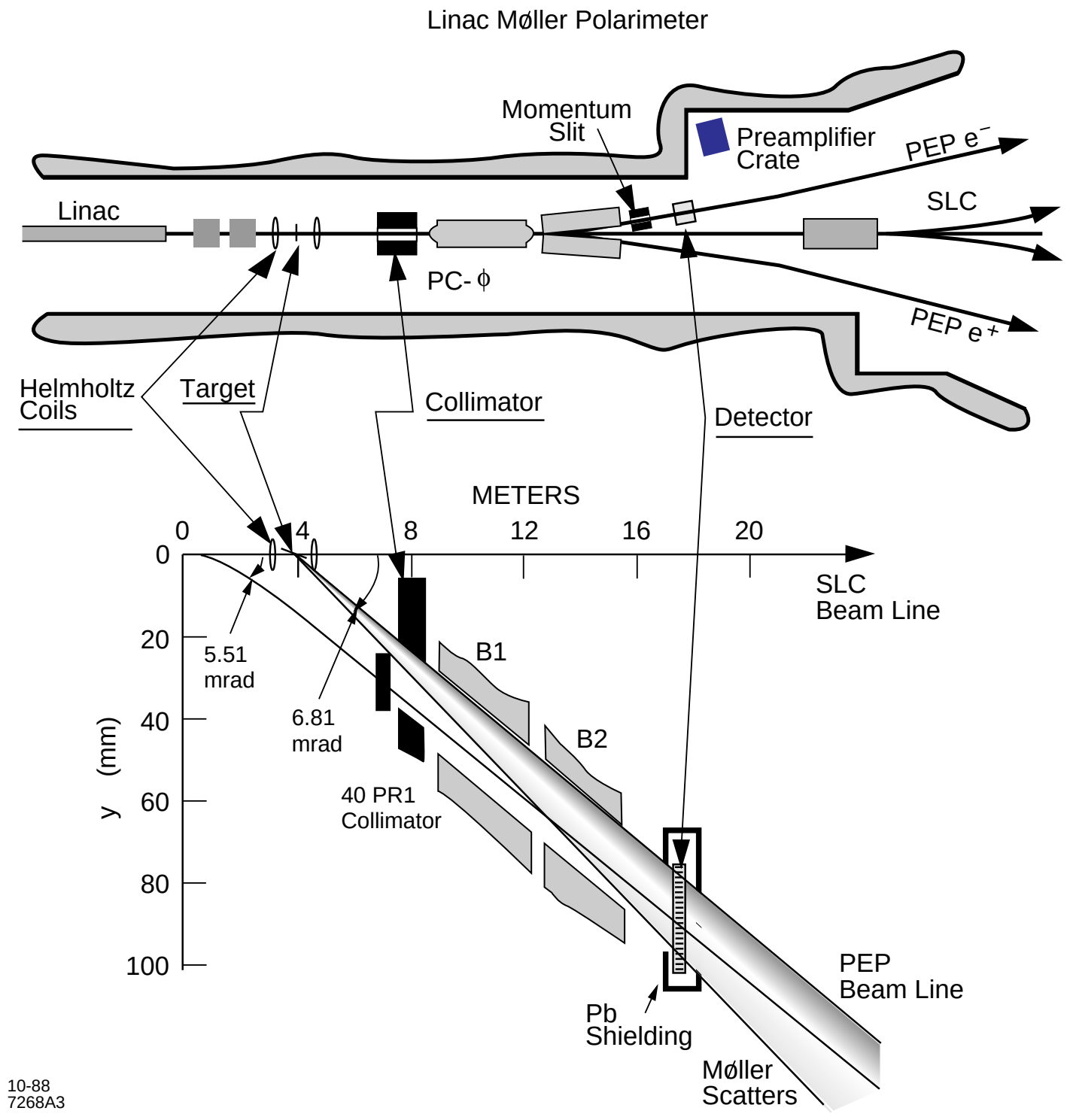

Figure 6. A schematic diagram of the SLC Linac Møller Polarimeter.

The Møller target assembly includes five Vanadium-Permendur foils which are mounted on a machined aluminum target holder. The work presented here makes use of two 9.5×133 mm longitudinal foils which are tilted by 20° with respect to the beam axis and have thicknesses 49 $\mu$m and 156 $\mu$m, respectively. A pair of Helmholtz coils generates a 120 Gauss magnetic field along the beam axis to magnetically saturate the target foils. The magnetization densities of the foils are determined from the difference of flux measurements performed with and without the targets present. The magnetization densities are corrected for the orbital contributions[8] to extract the target polarizations. The measured polarizations of the 49 $\mu$m and 156 $\mu$m foils are $0.0828 \pm 0.0027$ and $0.0790 \pm 0.0015$, respectively.[9]

The Lecroy 2259B peak sensing ADC that was used to digitize the amplified detector signals was found to have serious non-linearities in the lowest 10% of its 2 V input range. These were moderated somewhat by increasing the pedestal levels to approximately 300 counts (of the 2020-count full scale). The digitized signals were typically 50-300 ADC counts above the new pedestal. In this region, the electronic response functions (amplifier and ADC) deviate from an offset linear function by less than 3%. The deviations are corrected using a 16-segment piecewise linear function for each of the 64 amplifier/ADC channels.

The SLC Møller polarimeter is designed to operate at a center-of-mass scattering angle of 112° where the tree-level longitudinal Møller scattering asymmetry is 0.62. This operating point has somewhat less analyzing power than the commonly-used 90° point, but features less background contamination from radiative nuclear scattering. A beam pulse of $2 \times 10^{10}$ electrons incident upon the 49 $\mu$m target produces about 10 detected electrons. The analysis procedure is described in detail in Section 5.

## 4. The Experimental Procedure

The investigation of the Levchuk Effect makes use of eight data sets that were collected with the Linac Møller polarimeter in 1993. Two of these sets were collected as part of a program to determine the effect of the SLC arc transport system upon the polarization at the Compton Polarimeter. On those occasions (described below), it was possible to accurately determine the polarization in the linac from measurements made with the Compton polarimeter. Since the beam polarization measured at the Compton device was stable throughout the period during which the eight sets were collected, the two determinations of the beam polarization in the linac can be applied to the entire eight-set sample of Møller measurements.

### 4.1 Spin Transport Studies in the SLC North Arc

The transport of the electron beam through the SLC North Arc rotates the spin vectors of individual beam particles according to their energies. The finite energy width of the SLC electron beam (~0.2% RMS) implies that orientations of the spin vectors at the Compton polarimeter are distributed about the mean direction with a finite angular width. The net beam polarization measured at the Compton polarimeter is therefore less than the beam polarization in the linac.

The net arc spin rotation and polarization loss are measured according to the following procedure. The RMS energy width of a low current beam is reduced to less than 0.1%. This is accomplished by launching a shorter-than-normal electron bunch from the damping ring at an optimal (for energy width) RF phase in the linac. The resulting beam energy distribution is measured at a point of large energy dispersion in the SLC final focus region by passing a thin wire through the beam and observing the scattered radiation. The optimal spin orientation in the linac is then determined from longitudinal polarization measurements made with the Compton polarimeter for three non-planar linac polarization orientations. This procedure determines the coefficients, $a_x$, $a_y$, and $a_z$, which relate the longitudinal polarization at the Compton polarimeter to the linac polarization vector $\vec{\mathcal{P}}^L$,

$$\mathcal{P}_z^C = a_x \mathcal{P}_x^L + a_y \mathcal{P}_y^L + a_z \mathcal{P}_z^L. \qquad (10)$$

The linac spin direction given by the vector $(a_x, a_y, a_z)$ optimizes the longitudinal polarization at the Compton polarimeter. The spin rotation solenoids in the RTL and linac are then adjusted to launch the optimal spin orientation into the arc.

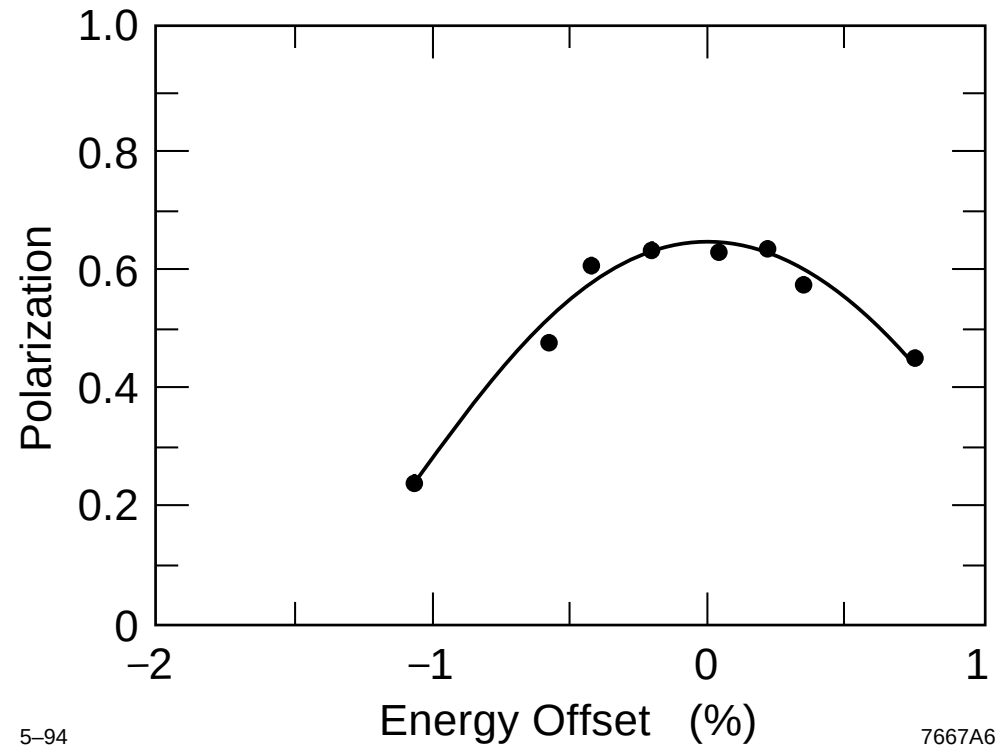

Figure 7. The measured energy dependence of the longitudinal beam polarization at the Compton polarimeter.

Finally, the beam energy ($E$) is varied in several steps by $\pm 0.9\%$ about the nominal 46.6 GeV arc launch energy ($E_0$) and the longitudinal polarization at the Compton polarimeter is measured at each energy.

The two sets of arc spin transport measurements give consistent results. In both cases, the optimal spin launch direction is found to be very close to the (nominal) vertical launch direction. The measured energy dependence of the longitudinal polarization at the Compton polarimeter is shown in Figure 7. The data are well-described by the following expression (a simple plane rotator model),

$$\mathcal{P}_z^C = \mathcal{P}_0 \cos\left[2\pi N_{eff}\left(\frac{E - E_0}{E_0}\right)\right], \tag{11}$$

where $\mathcal{P}_0$ is the peak polarization, and $N_{eff}$ is the effective number of spin precessions in the SLC arc which is found to be 17.9±0.2 from a fit to the data.

Using equation (11) and the measured beam energy distribution, the polarization values measured with the optimal launch direction are corrected by a factor of 1.006±0.002 to account for residual energy-spread-induced depolarization. An additional correction factor of 1.004±0.004 is applied to account for the randomization caused by synchrotron radiation as determined from Monte Carlo simulations. The net polarization in the linac is extrapolated to be

$$\mathcal{P}^L = 0.657 \pm 0.009,$$

where the error is the quadrature sum of the statistical and systematic uncertainties on the polarization measurements and the systematic uncertainty on the residual depolarization correction.

## 4.2 Møller Measurements

The Møller measurements are not compatible with normal SLC operation and require special running conditions. To make a measurement, the spin rotation system downstream of the SLC north damping ring is used to produce longitudinal polarization in the linac. A longitudinally polarized (20°) foil target is inserted into the linac and the unscattered beam is transported through the beam switchyard and part of the SLC North Arc to a beam dump.

Residual linear polarization of the polarized electron source laser beam can lead to small helicity-dependent beam current asymmetries. The net effect of these is minimized by reversing the polarization direction of the target foil between the 10-minute runs of the polarimeter. A typical measurement consists of four such runs. For each run, the total signal observed on each detector channel for both of the (randomly-changing) beam helicity states is recorded along with the total beam current for each helicity state and information on the polarimeter status.

Seven of the eight sets of data were taken with the 49 $\mu$m target and one set (set 4) with the 156 $\mu$m target. The beam energy and spectrometer setting were uniform for seven of the eight data sets. For these runs, the beam energy was 46.6 GeV, the polarization direction was aligned with the beam axis, and the central accepted momentum of the polarimeter was set to 14.5 GeV/c. The remaining data set (set 3) was measured with a 40.6 GeV beam energy and 14 GeV/c polarimeter setting. To further complicate matters, a problem with one of the spin rotation solenoids left the spin direction oriented at an angle of 49.8° with respect to the beam axis. The resulting transverse polarization component was in the vertical direction. Since the longitudinal target foils also have a vertical polarization component, the analysis of this run involves the longitudinal and transverse Møller asymmetry functions.

# 5. The Møller Analysis

## 5.1 The Monte Carlo Simulation

Since the actual signals observed in a single-arm Møller polarimeter depend strongly on a number of apparatus-dependent effects, we have performed a fairly complete Monte Carlo simulation of the Linac Møller Polarimeter. The initial position and angle coordinates of the interacting beam electrons are chosen from Gaussian distributions that have been scaled to model the beam emittance and the beta functions at the Møller target. The position and angle coordinates of the incident and scattered electrons are adjusted according to the Moliere parameterization for multiple Coulomb scattering in the target foils and vacuum window.[10] The energies of the incident and scattered electrons are adjusted to account for external bremsstrahlung in the target foils and vacuum window.[11] The detailed response of the detector package is simulated according to the parameterized results of a number of EGS4 simulations.[12]

The thicknesses of the target foils are less than or comparable to the equivalent radiator thickness for the *ee* scattering process at the SLC beam energy.[11] This implies that internal radiative processes are more important than the external radiative processes occurring in the target foils. Collinear initial and final state radiative effects are incorporated into the Monte Carlo simulation via the electron structure function approach. The resulting cross sections and asymmetries are checked against the complete first-order Monte Carlo calculation of Jadach and Ward.[13]

The simple collinear radiation model is based upon the approximation illustrated in Figure 8. In the center-of-mass frame of the beam and target electrons (the btcm-frame), the initial-state electrons can radiate the fractions $(1-x_1)$ and $(1-x_2)$ of their energies $\sqrt{s_1}/2$ ($s_1$ is defined in equation (7)) before colliding. Similarly, the detected final state electron can radiate the fraction $(1-x_3)$ of its energy into collinear photons. Photon emmission at finite angles and purely virtual corrections are neglected in this approximation. The tree-level differential cross section for polarized Møller scattering in the post-initial-state radiation center-of-mass (pisrcm) frame is given by equations (1) and (2) with $s$ replaced by $s_1x_1x_2$. The radiatively-corrected differential cross section is given by the product of the tree-level cross section and electron structure functions for each external leg of the process shown in Figure 8,

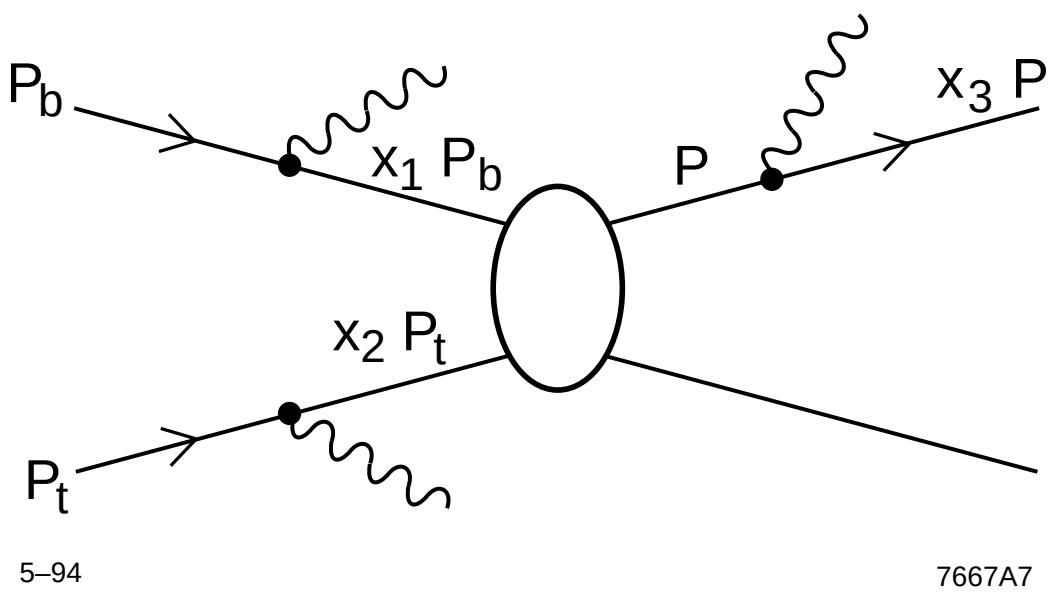

Figure 8. A diagram of the simple collinear radiation model used to simulate the effect of internal radiation upon the Møller scattering process.

$$\frac{d\sigma}{d\Omega dx_1 dx_2 dx_3} = \frac{d\sigma_0}{d\Omega}(s_1x_1x_2) \cdot D(x_1,T) \cdot D(x_2,T) \cdot D(x_3,T), \tag{12}$$

where the functions $D(x,T)$ are electron structure functions[14] at the momentum-transfer scale $T$. For this work, we assume that $T$ is the minimum of the magnitudes of the Mandelstam variables $|t_1|$ and $|u_1|$ defined in the absence of internal radiation,

$$T = \frac{s_1}{2}\left(1 - |\cos\hat{\theta}|\right).$$

The scattering angle and momentum of the final state electron in the laboratory frame are found by Lorentz boosting the pisrcm-frame momenta to the btcm-frame and then to the laboratory frame. The expressions given in equations (8) and (9) are modified as follows,

$$
\begin{aligned}
p' &= \frac{p_b x_1 x_3}{2}\left(1+\cos\hat{\theta}\right)\\
\theta^2 &= 2m_e x_2\left(\frac{x_3}{p'}-\frac{1}{p_b x_1}\right)\cdot\left(1-\frac{\vec{p}_t\cdot\hat{n}}{m_e}\right).
\end{aligned}
\tag{13}
$$

It is clear that internal radiation affects both the momentum and the angle of the scattered electron.

The simulation of the atomic momentum distributions for the target electrons is based upon screened hydrogen atom wavefunctions in momentum space. This approximation is reasonable for the K- and L-shell electrons which are bound to individual atomic sites. The outer-shell electrons in a metal form energy bands and are probably not described well by this approach. However, since most of the Levchuk line broadening is caused by the high-momentum, inner-shell electrons, an accurate description of the lower-momentum portion of the electron population is not necessary. The hydrogen atom wave functions[15] $\phi_{n\ell}(q)$ are normalized as follows,

$$
\int dq q^2 |\phi_{n\ell}(q)|^2 = 1, \tag{14}
$$

where: $q$ is the electron momentum in units of $Z\alpha m_e$ ($Z$ is the nuclear charge), $n$ is the principal quantum number, and $\ell$ is the angular momentum quantum number. The actual momentum distributions for unpolarized and polarized electrons, $f_{unp}(p)$ and $f_{pol}(p)$, are constructed as follows,

$$
\begin{aligned}
f_{unp}(p) &= \sum_{j,n,\ell}\frac{C^j_{n\ell}}{P^j_n}\left(\frac{p}{P^j_n}\right)^2 |\phi_{n\ell}(p/P^j_n)|^2\\
f_{pol}(p) &= \sum_{j}\frac{D^j_{32}}{P^j_3}\left(\frac{p}{P^j_3}\right)^2 |\phi_{32}(p/P^j_3)|^2,
\end{aligned}
\tag{15}
$$

where: $j$ labels the atomic species of the target foil, $C^j_{n\ell}$ is the fraction of the total unpolarized electron population that is associated with the $j^{th}$ species and the $n\ell$ orbital, $D_{32}^{j}$ is the fraction of the polarized d-wave, M-shell electrons that are associated with the $j^{th}$ species, and $P_n^j = Z_n^j \alpha m_e$ is an atomic momentum scale that has been adjusted to account for screening. The effective nuclear charge $Z_n^j$ is given by the simple ansatz that the nuclear charge seen by an electron is screened by all inner-shell electrons and one half of the same-shell neighbors,

$$Z_n^j = Z^j - \sum_i^{n-1} N_i^j - \frac{N_n^j - 1}{2}, \tag{16}$$

where $Z^j$ is the nuclear charge of the $j^{th}$ species and $N_i^j$ is the number of electrons in the $i^{th}$ shell.

The modelled K-, L-, M-, and N-shell momentum distributions for the iron atom are shown as solid curves in Figure 9. Note that the vertical scale is logarithmic. They are compared with the semiempirical K- and L-shell parameterizations of Chen, Kwei, and Tung[16] which are shown as boxes and crosses, respectively. The agreement is perfect for the K-shell distributions. The L-shell distributions agree well except at the largest momenta. The higher momentum shells have been compared with the Hartree-Fock calculations of Weiss, Harvey, and Phillips[17] and are found to agree well.

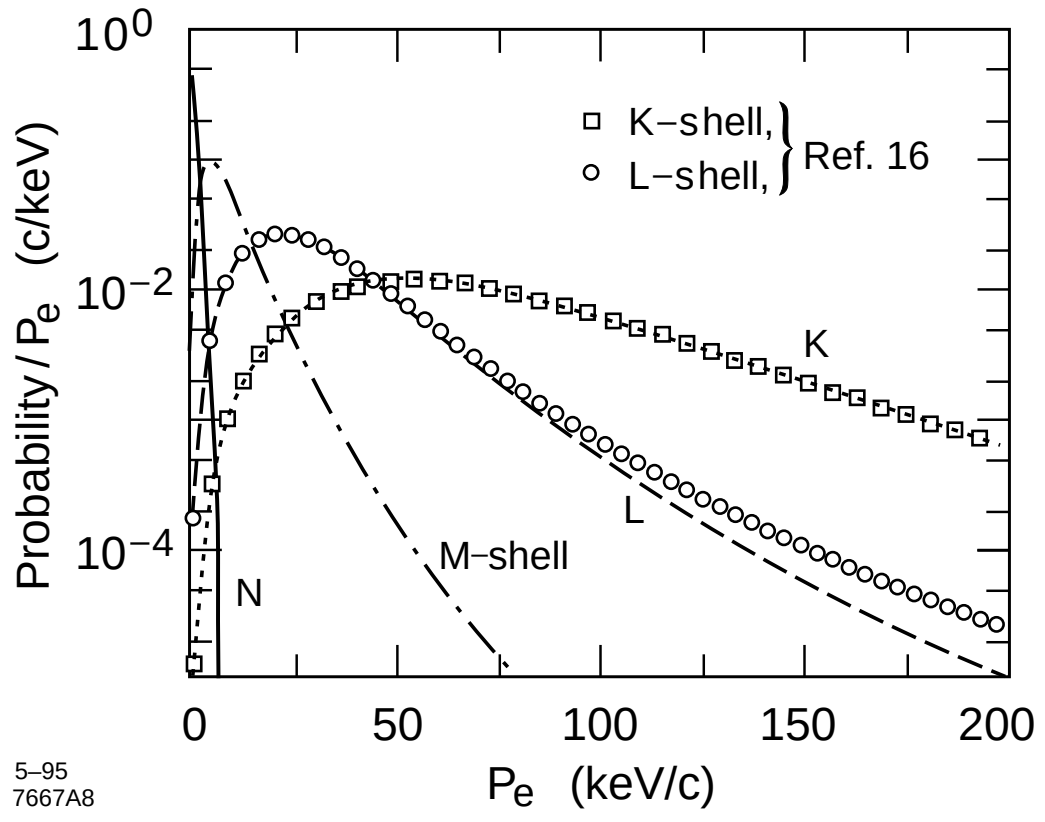

Figure 9. The modelled K-, L-, M-, and N-shell momentum distributions for the iron atom.

The simulated signal observed at the Møller detector per target electron is shown in Figure 2 for each of the atomic iron shells shown in Figure 9. Note that the peaks in Figure 2 associated with the K- and L- shell targets are substantially broadened and produce much less signal at the center of the distribution than do the M- and N-shell signals. This is a graphic illustration of the Levchuk Effect. The more highly polarized M-shell produces a larger Møller scattering asymmetry near the center of the peak. The asymmetry of the adjacent regions is diluted by the same effect and the overall width of the elastic peak is broadened.

The complete simulation is shown in Figure 10. The signal $S(y)$ and longitudinal scattering asymmetry $\mathcal{A}_z(y)$ are shown as functions of position $y$ on the detector. The solid curves incorporate all effects including the atomic momentum distributions (the wiggles in the asymmetry function are caused by limited Monte Carlo statistics in regions of small accepted cross section). The dashed curves show the same simulation with zero atomic momenta. Note that the asymmetry function (analyzing power) is increased by 14% at the Møller peak and is substantially diluted in the adjacent regions.

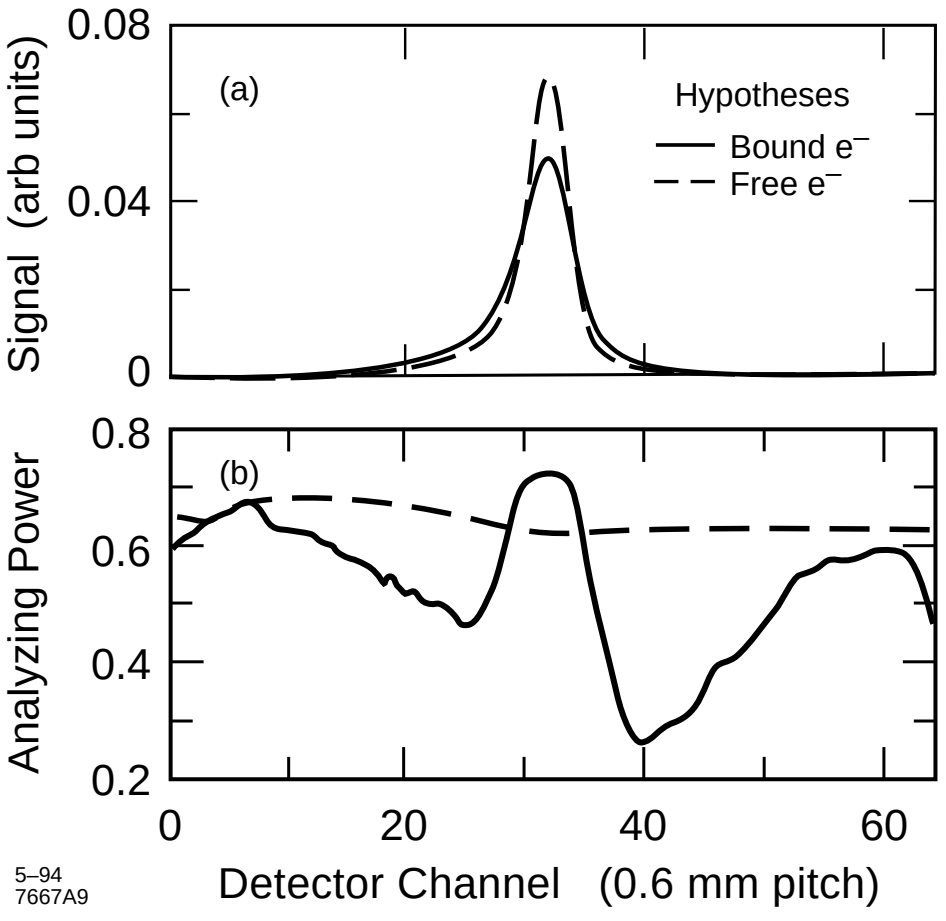

Figure 10. The complete simulation of the signal and longitudinal scattering asymmetry observed at the Møller detector.

### 5.2 The Fitting Procedure

The polarimeter functions by recording the average signal in each detector channel for the two beam helicity states. The target helicity is reversed on successive runs. The data for the four helicity combinations are combined into average signals for the case where the beam and target spins are antiparallel, $N(j, \lambda = -1)$, and parallel, $N(j, \lambda = 1)$, where $j$ labels the detector channels and $\lambda$ labels the relative beam-target helicity state. Combining the data in this manner suppresses the small helicity-dependent asymmetry in the electron current which can be produced by residual linear polarization in the electron source laser beam (typically $\lesssim 0.1\%$). The net beam current asymmetry $A_e$ is directly measured with toroid beam current monitors in the linac.

The detected signals are produced by a number of processes. The Møller scattering process produces high energy electrons which are directly accepted by the spectrometer but also shower on acceptance edges producing a diffuse signal at the detector. Nuclear scattering with internal or external radiation and several related processes can also produce high energy electrons which are accepted by the spectrometer. Finally, beam halo and target-related collision products can produce signal on the most well-shielded detectors. To account for these processes, the signals $N(j, \lambda)$ are fit simultaneously to the sum of the Møller signal shape derived from the Monte Carlo simulation and a smooth (non-peaked) empirical background function which can also depend upon $\lambda$ (to account for diffuse Møller scattering background). Another effect which occurs in the real polarimeter is that the vertical beam position can change from time to time. The Monte Carlo simulation shows that the measured signal shape and asymmetry function are insensitive to the small (<1 mm) changes but are translated by the beam motion. The fitting procedure therefore allows for translations of the detected signal. The actual fits are performed by minimizing the $\chi^2$ function which is defined as follows,

$$\chi^2 = \sum_{\lambda,j} \frac{\left[N(j,\lambda) - \mathcal{R}(y_j + \Delta, \lambda)\right]^2}{\sigma^2(j,\lambda)}, \tag{17}$$

where: $\sigma(j, \lambda)$ is the statistical uncertainty on $N(j, \lambda)$ ; $y_j$ is the position of the $j^{th}$ channel, and $\Delta$ is a parameter to translate the fitting function $\mathcal{R}(y, \lambda)$. The fitting function is defined as follows,

$$\begin{aligned} \mathcal{R}(y,\lambda) = (1 - \lambda A_e) \Bigg\{ & R_N S(y) \left(1 - \lambda \left[\mathcal{P}_z^L \mathcal{P}_z^T \mathcal{A}_z(y) + \mathcal{P}_y^L \mathcal{P}_y^T \mathcal{A}_t(y)\right]\right) \\ & + \sum_{i=0}^{n} (b_i - \lambda c_i)\, y^i \Bigg\}, \end{aligned} \tag{18}$$

where: $A_e$ is the measured beam current asymmetry; $R_N$ is a normalization parameter; $S(y)$, $\mathcal{A}_z(y)$, and $\mathcal{A}_t(y)$ are the signal and asymmetry functions derived from the Monte Carlo simulation; $\mathcal{P}_z^L$ and $\mathcal{P}_z^T$ are the longitudinal polarizations of the beam and target, respectively; $\mathcal{P}_y^L$ and $\mathcal{P}_y^T$ are the vertical polarizations of the beam and target, respectively; $b_i$ and $c_i$ are coefficients of the helicity-dependent polynomial background; and $n$ is the order of the background polynomial.

The Monte Carlo simulation does not include the aperture constraints caused by the downstream vacuum chamber. In the polarimeter data, small changes in the signal shape are observed near detector channels 16 and 48 indicating the onset of the vacuum chamber aperture constraints. This observation is supported by tests in which the accepted momentum was varied and the peak position moved into the obscured regions. The presence of downstream aperture restrictions explains why substantial non-zero asymmetry was observed in the wings of the distribution. The obscured regions are removed from the analysis by restricting the fits to the detector channels $j$ where $17 \leq j \leq 48$.

## 5.3 Systematic Uncertainties

The systematic uncertainties associated with the Møller polarimeter measurements are summarized in Table 2. The measurements of the target foil polarizations are uncertain at the $\pm 3.1\%$ level and lead to a $\pm 3.1\%$ fractional uncertainty on the measured beam polarization. The measured beam polarization is slightly sensitive to the order of the background polynomial. Changing the order of the background polynomial used in the fitting procedure from one to five causes the beam polarization estimate $\mathcal{P}^L$ to vary by no more than 2.1% of itself. We take this value as a conservative estimate of the uncertainty associated with the background parameterization. The corrections for the response functions of the detector preamplifiers and ADC system modify $\mathcal{P}^L$ by 3.6% of itself (they also decrease the average fit $\chi^2$ by a factor of 1.7). The uncertainty associated with these corrections is estimated to be $\pm 1\%$. The momentum scale of the polarimeter is determined from the position of the two-body elastic peak on the detector (and the measured detector position). The uncertainty on the momentum scale is 1.8% which leads to a $\pm 1.4\%$ uncertainty on $\mathcal{P}^L$.

There are several uncertainties associated with the Monte Carlo model. The sensitivity of the result to the simulated atomic momentum distributions is inferred by repeating the analysis with the delta function distributions used by Levchuk in Reference 1. Although this causes the fit quality to be degraded somewhat (the $\chi^2$ values are increased by an average factor of 1.4), the fractional change in the beam polarization is smaller than 0.2%. Varying the bremsstrahlung and multiple scattering parameterizations produces slightly larger effects. The radiative corrections used in the Monte Carlo simulation are based upon the simple collinear radiation model which ignores radiation at finite angles and purely virtual corrections. We estimate the size of the omitted effects by comparing our simulated results with those obtained from the Monte Carlo generator of Jadach and Ward.[13] The two calculations deviate by less than 0.5%. The overall modelling uncertainty is conservatively estimated to be ±1%.

The overall systematic uncertainty on the polarization scale is ±4.2%.

**Table 2**. Systematic uncertainties that affect the Linac Møller polarimeter measurements.

| Systematic Uncertainty | $\delta\mathcal{P}^L/\mathcal{P}^L$ |
|---|---|
| Target Polarization | 3.1% |
| Background Parameterization | 2.1% |
| Electronic Response Corrections | 1.0% |
| Spectrometer Momentum Scale | 1.4% |
| Modelling Uncertainties | 1.0% |
| Total Uncertainty | 4.2% |

### 5.4 Results

The fitting procedure described in Section 5.2 was applied to the eight sets of data taken during the summer of 1993. All results presented in this section are based upon a linear background polynomial ($n = 1$). Two atomic momentum hypotheses were used to simulate the signal and asymmetry functions. The first hypothesis assumes that the target electrons are at rest and is labelled *free-electron-target*. The second hypothesis uses the atomic momentum distributions and is labelled *bound-electron-target*.

Typical fits of these hypotheses to a single set of data (set 5) are shown in Figures 11 and 12. The signal and asymmetry measured by each detector channel are plotted as solid points. The statistical uncertainty on each signal measurement is much smaller than the point size (typically 0.1% of the signal size). The fits are shown as solid histograms. The dashed lines indicate the size of the background signal and asymmetry. The free-electron-target hypothesis clearly underestimates the observed width of the signal and yields the polarization measurement, $\mathcal{P}^L = 0.824 \pm 0.027$, where the error is entirely statistical. The bound-electron-target hypothesis provides a much better estimate of the signal shape and yields the polarization measurement, $\mathcal{P}^L = 0.705 \pm 0.024$.

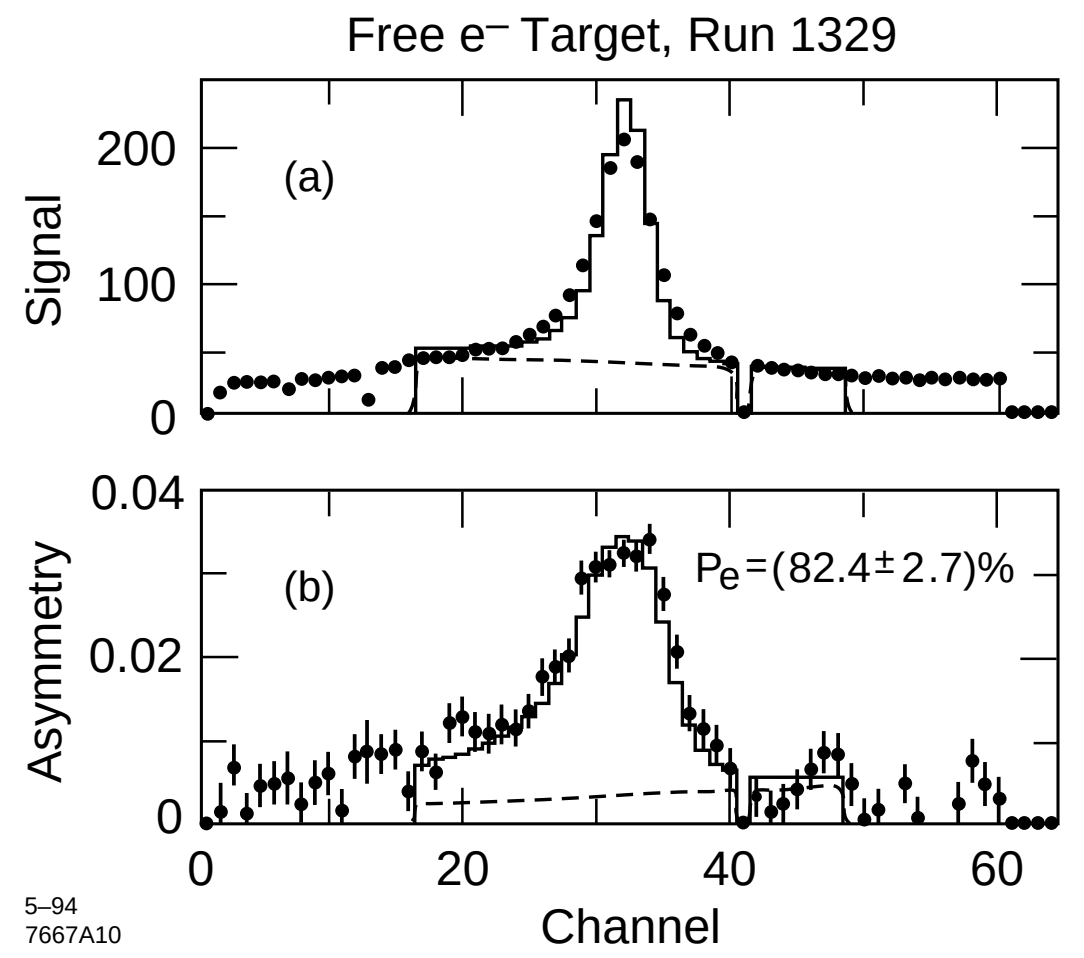

Figure 11. The measured signal and asymmetry for data set 5 are plotted as solid points. The signal errors are much smaller than the point size. The best fit to the free-electron-target hypothesis is shown as a solid histogram. The dashed line indicates the background signal size and asymmetry.

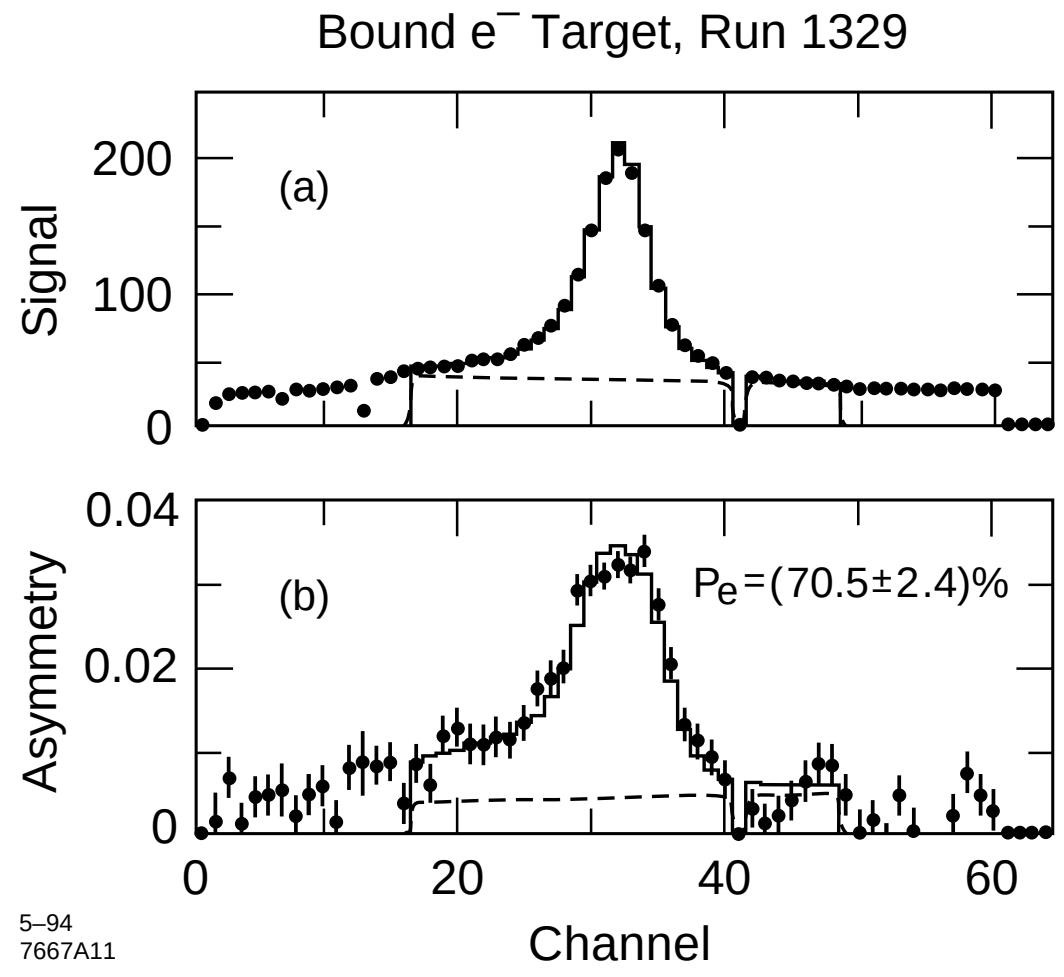

Figure 12. The measured signal and asymmetry for data set 5 are plotted as solid points. The signal errors are much smaller than the point size. The best fit to the bound-electron-target hypothesis is shown as a solid histogram. The dashed line indicates the background signal size and asymmetry.

The results of fitting all eight data sets are summarized in Figure 13. The beam polarization estimates derived from the free-electron-hypothesis are plotted as diamonds and those derived from the bound-electron-hypothesis are plotted as squares. The plotted error bars reflect the statistical uncertainties only. Note that the third measurement that was made at a non-standard beam energy and spin orientation is consistent with the others. The mean free-electron-target and bound-electron-target results,

$$\bar{\mathcal{P}}^L = \begin{cases} 0.800 \pm 0.009(\text{stat.}) \pm 0.034(\text{syst.}), & \text{free-electron-target hypothesis} \\ 0.690 \pm 0.008(\text{stat.}) \pm 0.029(\text{syst.}), & \text{bound-electron-target hypothesis,} \end{cases}$$

are plotted at the right of the figure and include the systematic errors. The linac polarization as determined from the Compton measurements ($0.657 \pm 0.009$) is also shown in Figure 13 and is 1.1 standard deviations smaller than the bound-electron-target result. The free-electron-target result deviates from the Compton result by 4.1 standard deviations.

Further support for the bound-electron-target hypothesis comes from examining the goodness-of-fit parameter $\chi^2$ for the two hypotheses. Like most polarimeter results, the $\chi^2$ values associated with both hypotheses is poor. This is a consequence of the enormous statistical precision of the signal measurements ($\lesssim 0.1\%$) and the impossibility of gain-matching the channels and calculating the signal shape to the same level of precision. Nevertheless, we can compare the hypotheses by considering the ratio of the $\chi^2$ values associated with the two fit hypotheses (the ratios are so large that the more traditional difference of $\chi^2$ isn't meaningful). The ratio of the $\chi^2$ for the bound-electron-target hypothesis to that for the free-electron-target hypothesis for each data sample is shown in the lower plot of Figure 13. The mean ratio, 0.083, is shown as the solid horizontal line. It is clear that the bound-electron-target hypothesis is strongly favored.

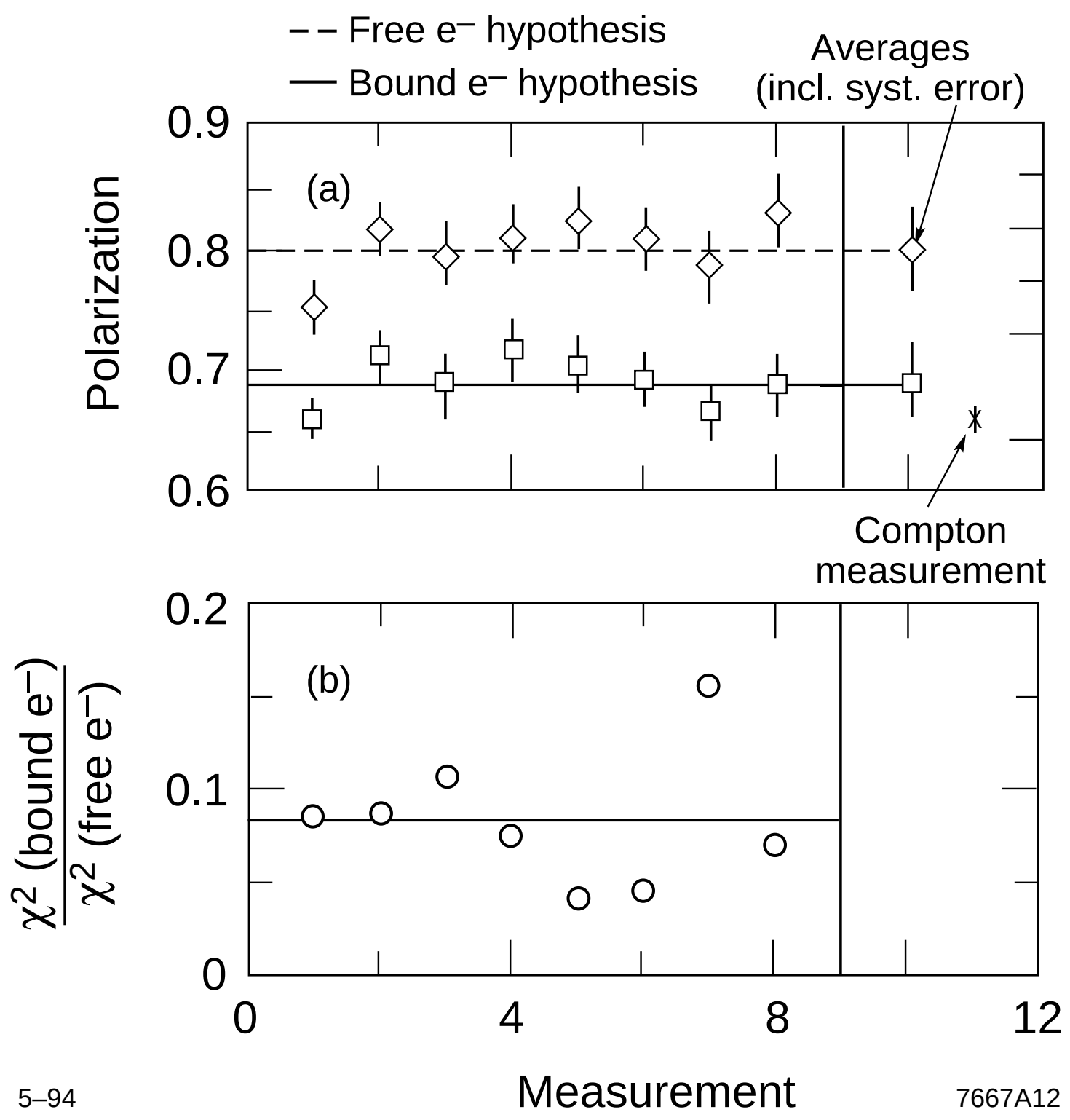

Figure 13. The results of fitting the free-electron-target and bound-electron-target hypotheses to the eight data samples.

## 6. Conclusions

The traditional approach to the analysis of data from a single-arm Møller polarimeter is to empirically parameterize the measured shapes of the two-body elastic peak and the background distribution. These shapes are used to infer the signal-to-background ratio. This approach is based upon the assumption that the asymmetry function is uniform across the detector image. In this paper, we have shown that this assumption is false. The presence of non-zero target electron momenta can cause significant non-uniformities in the asymmetry function. The same effect significantly broadens the elastic peak and must be incorporated into a simulation of the lineshape. The resulting lineshape calculation has the advantage that it reduces the number of free parameters in the fitting function and provides a more reliable background estimate.

The Levchuk Effect has been observed with the SLC Linac Møller polarimeter. The effect alters the measured beam polarization by 14% of itself and must be corrected to achieve consistency with beam polarization measurements performed with a precise Compton polarimeter. Additionally, the effect is needed to describe the measured width of the elastic peak. The correction to the measured polarization is not universal but depends upon the details of the polarimeter construction, beam parameters, and analysis technique. The non-universality of the correction makes it difficult to estimate the impact of the Levchuk Effect upon physical measurements performed in the past with single-arm Møller polarimeters. The estimation of corrections requires detailed information about each specific polarimeter and analysis.

Finally, we note that this paper has been primarily addressed to single-arm polarimeters. That is because the operation of single-arm devices requires high angular resolution to separate signal and background. Many double-arm Møller polarimeters are currently in use around the world. Since these devices use timing to identify the signal, they frequently have large acceptance and poor resolution in laboratory scattering angle. They are therefore less likely to be seriously affected by the Levchuk Effect. Nevertheless, it is not possible to globally rule-out the consequences of non-zero target electron momenta. As with single-arm devices, each individual case must be evaluated in detail.

Acknowledgements:

We would like to thank B.F.L. Ward and S. Jadach for making a preliminary version of their radiative Møller scattering Monte Carlo available to us. This work was supported in part by Department of Energy Contract No. DE-AC03-76SF00515.